\documentclass[pra,aps,reprint,superbib,superscriptaddress,twocolumn]{revtex4-1}

\usepackage{amsmath,bm}
\usepackage{amstext}
\usepackage{epsfig}
\usepackage{xcolor}
\usepackage{subfig}
\usepackage{graphicx}
\usepackage{multirow}
\usepackage{array}
\usepackage{tikz,pgfplots}
\usepackage{MnSymbol}
\definecolor{dgreen}{rgb}{0,.5,0}
\definecolor{dred}{rgb}{.7,.0,.0}

\newcommand{\etal}{{\it et al.}}

\def\ddroit{{\rm d}}
\newcommand{\be}{\begin{eqnarray}}
\newcommand{\ee}{\end{eqnarray}}

\begin{document}

\title{
Ground and excited energy levels can be extracted exactly
from a single ensemble density-functional theory calculation 
}

\author{Killian Deur}
\affiliation{Laboratoire de Chimie Quantique,
Institut de Chimie, CNRS/Universit\'e de Strasbourg,
4 rue Blaise Pascal, 67000 Strasbourg, France}

\author{Emmanuel Fromager}
\thanks{Corresponding author}
\email{fromagere@unistra.fr}
\affiliation{Laboratoire de Chimie Quantique,
Institut de Chimie, CNRS/Universit\'e de Strasbourg,
4 rue Blaise Pascal, 67000 Strasbourg, France}


\begin{abstract}

Gross--Oliveira--Kohn density-functional theory (GOK-DFT) for ensembles is the
DFT analog of state-averaged wavefunction-based (SA-WF) methods. In
GOK-DFT, the state-averaged (so-called ensemble) exchange-correlation
(xc) energy
is described by a single functional of the density which, for a fixed
density, depends on the weights assigned to each state in the ensemble.
We show that, if a many-weight-dependent xc functional is employed, then it becomes
possible to extract, in principle exactly, all individual energy levels
from a single GOK-DFT calculation, exactly like in a SA-WF calculation.
More precisely, starting from the Kohn--Sham energies, a global
Levy--Zahariev-type shift as well as a state-specific (ensemble-based)
xc derivative correction must be applied in order to reach the energy
level of interest. We illustrate with the asymmetric Hubbard
dimer the importance and substantial weight dependence of both
corrections. A comparison with more standard extraction procedures, which
rely on a sequence of ensemble calculations, is made at the
ensemble exact exchange level of approximation.
\end{abstract}

\maketitle

\section{Introduction}

Time-dependent density-functional theory (TD-DFT)~\cite{runge1984density} has become over the last two
decades the method of choice for modeling excited-state
properties~\cite{Casida_tddft_review_2012}.
Despite this success, it still suffers, in its standard (adiabatic)
formulation, from various limitations. The absence of
multiple-electron-excitation energies in the spectrum is one well-known
example~\cite{Casida_tddft_review_2012}. Moreover, as it relies on a
ground-state DFT calculation, linear response
TD-DFT does not provide a balanced description of
low-lying excited states. Such a description is of primary importance in
photochemistry when approching,
for example, an avoided crossing or a conical intersection, but also for
modeling the electronic structure of open $d$- or $f$-shell systems.\\

One way to overcome these limitations is to extend DFT to (canonical)
ensembles of ground and excited
states~\cite{PRA_GOK_EKSDFT,GOK3}.
Ensemble DFT relies on the Gross--Oliveira--Kohn (GOK)
variational principle~\cite{PRA_GOK_RRprinc}, which is a generalization
of Theophilou's variational principle for
equi-ensembles~\cite{JPC79_Theophilou_equi-ensembles,theophilou_book},
hence the name GOK-DFT.
Even though it is rarely mentioned, theses principles
provide a rigorous justification for the state-averaging
procedure that is routinely used in 
complete active space
self-consistent field (CASSCF) calculations~\cite{mcscf_pinkbook}.
GOK-DFT has been formulated thirty years ago and, despite important conceptual
progress~\cite{Nagy_ensAC,ensemble_ghost_interaction}, it did not attract as much attention as TD-DFT until now.
Quite recently, numerous important contributions (both formal and
practical) appeared in the
literature~\cite{PRA13_Pernal_srEDFT,franck2014generalised,yang2014exact,JCP14_Burke_ensemble,pernal2015excitation,yang2017direct,gould2017hartree,gould2018charge,deur2017exact,deur2018exploring,arXiv18_Gould_ens_corr_ener_not_just_sum,sagredo2018can,senjean2018unified,JCP14_Filatov_conical_inter_REKS,Filatov-2015-Wiley,filatov2015ensemble,filatov2016self,filatov2017description}, thus making GOK-DFT an active
field of research and a promising
time-independent alternative to TD-DFT.\\

Modeling the correlation energy of an ensemble with a density functional
is a complicated task since it is not, in general,  a simple sum
of individual correlation
energies~\cite{arXiv18_Gould_ens_corr_ener_not_just_sum}. Extracting 
individual energy levels is therefore not straightforward in
GOK-DFT~\cite{senjean2015linear,yang2017direct}. In the state-averaged CASSCF method the situation is different since the
(wavefunction-based) energy of each state is always computed, thus
giving access to excited-state properties (like energy gradients). From
that point of view, a state-specific DFT~\cite{PRA12_Nagy_TinD-DFT_ES,JCP15_Ayers_KS-DFT_excit-states_Coulomb,Ayers2018_Article_Time-independentDensityFunctio}
might be more appropriate.
Nevertheless, as mentioned
previously, it is often important, for example in photochemistry, to
have a balanced description (in terms of orbitals) of
ground and lower excited states. In such cases, using the ensemble formalism is
clearly relevant. Surprizingly, the flexibility of the theory regarding
the choice of the ensemble weights~\cite{PRA_GOK_RRprinc} has not been
fully explored yet. In standard GOK-DFT-based methods,
excitation energies are usually extracted from a sequence of ensemble
calculations (each of them involving a {\it single} ensemble
weight)~\cite{PRA_GOK_EKSDFT,senjean2015linear,yang2017direct}.
The Kohn--Sham DFT
limit (where all the excited-state weights become zero)
has been explored in this context, thus leading to the {\it direct ensemble
correction} (DEC) scheme of Yang
\etal~\cite{yang2017direct,sagredo2018can}. In this paper, we explore an
alternative formulation of GOK-DFT where a single {\it many}-weight-dependent ensemble exchange-correlation (xc)
functional is employed. In this formalism, all the weights can vary
independently. We show that, with such a flexibility, all individual
energy levels can be extracted, in principle exactly, from a single
GOK-DFT calculation where the ensemble weights can be freely chosen.
In contrast to TD-DFT, which gives access to excitation
energies only, this {many}-weight-dependent
formulation of GOK-DFT provides {\it total} excited-state energies.
Therefore, it should allow for a direct calculation of excited-state
properties by differentiation of the latter energies with respect to any perturbation
strength (like the nuclear displacements for the optimization of
equilibrium structures, for example).
We show that our {many}-weight-dependent approach is nothing but a generalization of DEC to
non-zero weights. As a result, it allows for a balanced description of the 
states within the ensemble through the adjustment of the
weights, exactly like
in a state-averaged CASSCF calculation.\\

The paper is organized as follows. After a brief review of the GOK
principle and the various extraction procedures of individual energy levels from
an ensemble calculation (Sec.~\ref{subsec:Ens_ener_deriv_ind_ener}), we
derive in Sec.~\ref{subsec:ind_ener_gok-dft} a {\it
many}-weight-dependent version of GOK-DFT where all the energy levels
can be determined from a single calculation. The connection with
existing ensemble DFT methods is made in Sec.~\ref{subsec:connection-standard}. The theory is then applied to
the asymmetric Hubbard dimer in Sec.~\ref{sec:Hubb_dimer_theory}. The
results are discussed in Sec.~\ref{sec:results-perspec}. Comparison is
then made, at the ensemble exact exchange level of
approximation,
with the more standard extraction technique, where a
sequence of ensemble calculations is performed (see
Sec.~\ref{sec:compar_gokII}). Conclusions and perspectives are given in Sec.~\ref{conclusions}.

\section{Theory}

\subsection{Extracting individual energy levels from an ensemble
energy}\label{subsec:Ens_ener_deriv_ind_ener}

Let us consider a canonical ensemble consisting of the ground and $M$
first
excited states of the electronic Hamiltonian $\hat{H}=\hat{T}
+\hat{W}_{\rm ee}+\hat{V}_{\rm ext}$. The operators $\hat{T}$ and $\hat{W}_{\rm
ee}$ describe the electronic kinetic and repulsion
energies, respectively. The local external potential operator reads $\hat{V}_{\rm ext}=\int\ddroit\mathbf{r}\,v_{\rm
ext}(\mathbf{r})\hat{n}(\mathbf{r})$ where $\hat{n}(\mathbf{r})$ is the
density operator and $v_{\rm
ext}(\mathbf{r})$ will simply be the nuclear Coulomb potential in this
work.   
For the sake of clarity, we will assume in the following that none of
these states are degenerate. The formalism can be easily extended to
degenerate ensembles by assigning the same weight to degenerate
states~\cite{PRA_GOK_EKSDFT,yang2017direct}. In the most general
formulation of the GOK variational principle~\cite{PRA_GOK_RRprinc}, the exact ensemble energy reads
\be\label{eq:ens_ener}
E^{\mathbf{w}}=\left(1-\sum^M_{I=1}{\tt w}_I\right)E_0+\sum^M_{I=1}{\tt
w}_IE_I,
\ee     
where $E_0$ is the ground-state energy, $\left\{E_I\right\}_{1\leq I\leq
M}$ are
the $M$ first excited-state energies, and $\mathbf{w}\equiv ({\tt
w}_1,{\tt
w}_2,\ldots,{\tt
w}_M)$ denotes the collection of weights that are assigned to each
individual {\it excited} state. In their seminal
paper~\cite{PRA_GOK_EKSDFT}, Gross {\etal} considered a
{\it sequence} of ensemble DFT calculations in order to extract excitation
energies. In their approach, each (non-degenerate here) ensemble is a linear interpolation (controlled by a {\it single}
ensemble weight $w$) between equi-ensembles:
\be\label{eq:gokI_weights}
{\tt w}_{1\leq I<M}=\dfrac{1-w}{M},\hspace{0.2cm}{\tt w}_M=w.
\ee
More recently, Yang {\it et al.}~\cite{yang2017direct} used another set
of ensembles (the approach was referred
to as GOKII) which are
also characterized by a {single} weight $w$:
\be\label{eq:gokII_weights}
{\tt w}_{1\leq I\leq M}=w.
\ee
The practical advantage of Eq.~(\ref{eq:gokII_weights}) over
Eq.~(\ref{eq:gokI_weights}) is that two ensemble calculations are
sufficient for extracting any excitation energy~\cite{yang2017direct}.
In Ref.~\cite{yang2017direct}, the authors implemented
Eq.~(\ref{eq:gokII_weights}) in the $w\rightarrow 0$ limit, thus
providing a {\it direct ensemble
correction} (DEC) to Kohn--Sham (KS) excitation energies.\\ 

One practical drawback of both DEC and linear response TD-DFT
is that, in contrast to state-averaged CASSCF~\cite{mcscf_pinkbook}, it
is not straightforward to study, within their formalisms, the potential
energy
curve of one or more excited states, simply because a sequence of different
calculations is
needed. Moreover (and perhaps, more importantly) none of them provides a balanced description (in
terms of orbitals) of the
ground and lower excited states. This can become problematic, for example, in the
vicinity of a conical intersection.\\

In order to address these deficiencies, we explore in this paper a more general formulation of GOK-DFT     
where the ensemble weights can all vary {\it independently}. Note that the ensemble energy can be obtained
variationally if the weights decrease with increasing
index~\cite{PRA_GOK_RRprinc}, i.e. if, for 
$1\leq J\leq (M-1)$,
\be\label{eq:convex_weights_1}
{\tt w}_J\geq {\tt w}_{J+1}\geq0,
\ee
and
\be\label{eq:convex_weights_0}
\left(1-\sum^M_{I=1}{\tt w}_I\right)\geq{\tt w}_1.
\ee   
Before introducing our alternative extraction procedure, we would like
to stress that, unlike state-averaged wavefunction-based methods,
GOK-DFT gives a direct access to the ensemble energy
$E^{\mathbf{w}}$ only, and not to its individual-state components (i.e. the
energy levels). The reason
is that, in GOK-DFT, a single density functional is used for describing the
xc energy of the ensemble. In the latter are mixed, in a non-trivial
way, the
individual correlation energies of all the states that belong to the
ensemble~\cite{arXiv18_Gould_ens_corr_ener_not_just_sum}.\\
 
Even though
excitation (or individual) energies cannot be extracted from a single
ensemble energy value $E^{\mathbf{w}}$, infinitesimal variations in the ensemble
weights will immediately give access to its individual components.
Indeed, starting from the fact that the derivative of the ensemble
energy with respect to ${\tt w}_I$ is equal to the $I$th excitation energy,
\be\label{eq:Xener}
\dfrac{\partial E^{\mathbf{w}}}{\partial {\tt w}_I}=E_I-E_0,
\ee
and keeping in mind that the ensemble energy varies {\it linearly} with the
ensemble weights (see Eq.~(\ref{eq:ens_ener})), 
\be
E^{\mathbf{w}}=E_0+\sum^M_{I=1}{\tt w}_I\dfrac{\partial E^{\mathbf{w}}}{\partial
{\tt w}_I},
\ee   
or, equivalently,
\be\label{eq:E0_wrt_ens_ener}
E_0=
E^{\mathbf{w}}
-\sum^M_{I=1}{\tt w}_I\dfrac{\partial E^{\mathbf{w}}}{\partial
{\tt w}_I},
\ee   
we can rewrite any individual (ground- or excited-state) energy
as
\be\label{eq:ind_ener_exp_deriv_wft}
E_K&=&E_0+\sum^M_{I=1}\delta_{IK}(E_I-E_0)
\nonumber\\
&=&E^{\mathbf{w}}+\sum^M_{I=1}\left(\delta_{IK}-{\tt
w}_I\right)\dfrac{\partial E^{\mathbf{w}}}{\partial {\tt w}_I},
\ee
where $0\leq K\leq M$. 
The derivation of
Eq.~(\ref{eq:ind_ener_exp_deriv_wft}) is trivial. Nevertheless, to the best
of our knowledge, it has never been
used in the context of GOK-DFT. As shown in the following, the 
expression in Eq.~(\ref{eq:ind_ener_exp_deriv_wft}) is convenient for connecting the
exact individual energy levels to the KS orbital
energies. Most importantly, it will
enable us to show that a single GOK-DFT calculation (where the
weights can be freely chosen) is in principle sufficient for extracting
all the energy
levels.

\subsection{Density-functional theory for
ensembles}\label{subsec:ind_ener_gok-dft}

In GOK-DFT, the ensemble energy is determined variationally as
follows~\cite{PRA_GOK_EKSDFT},
\begin{eqnarray}\label{eq:min_gamma_Ew}
 E^{\mathbf{w}} &=& \min_{{\hat{\gamma}}^{\mathbf{w}}}
\Big\{ 
{\rm Tr}\left[\hat{\gamma}^{\mathbf{w}}
\left(\hat{T}
+\hat{V}_{\rm ext}\right)
\right]+ E_{\rm
Hxc}^{\mathbf{w}}[n_{\hat{\gamma}^{\mathbf{w}}}]
\Big\}
\nonumber\\
&=&
{\rm Tr}\left[\hat{\gamma}_{\rm s}^{\mathbf{w}}
\left(\hat{T}
+\hat{V}_{\rm ext}\right)
\right]+ E_{\rm Hxc}^{\mathbf{w}}[n_{\hat{\gamma}_{\rm s}^{\mathbf{w}}}]
,
\end{eqnarray}
where $n_{\hat{\gamma}^{\mathbf{w}}}(\mathbf{r})={\rm Tr}\left[\hat{\gamma}^{\mathbf{w}}
\hat{n}(\mathbf{r})
\right]$ is a trial ensemble density and  
\be\label{eq:Hxc_ens_fun}
E^{\mathbf{w}}_{\rm Hxc}[n]
=
 \frac{1}{2} \iint
\ddroit{\bf r} \ddroit{\bf r'}
~ \dfrac{n({\bf r})n({\bf r'})}{\mid{{\bf r}-{\bf r'}}\mid}
+
E^{\mathbf{w}}_{\rm xc}[n]
\ee
is the ensemble Hartree xc (Hxc) functional. We use here the original
in-principle-exact decomposition of the Hxc functional~\cite{PRA_GOK_EKSDFT} where, for a given and fixed
density $n$, the xc part only varies with $\mathbf{w}$. In practical
(approximate) calculations, it might be worth using another
decomposition~\cite{arXiv18_Gould_ens_corr_ener_not_just_sum} which is
ghost-interaction-free~\cite{ensemble_ghost_interaction}. In this work, we will
always use exact Hxc (or Hx) functionals. Returning to Eq.~(\ref{eq:min_gamma_Ew}), the ground and excited KS determinants in the minimizing
non-interacting density matrix operator $\hat{\gamma}_{\rm
s}^{\mathbf{w}}=\left(1-\sum^M_{I=1}{\tt w}_I\right)
\left\vert\Phi^{\mathbf{w}}_0\right\rangle\left\langle\Phi^{\mathbf{w}}_0\right\vert
+
\sum^M_{I=1}{\tt
w}_I\left\vert\Phi^{\mathbf{w}}_I\right\rangle\left\langle\Phi^{\mathbf{w}}_I\right\vert
$ 
are determined by solving the ensemble KS equations self-consistently,
\be\label{eq:sc_ks-edft}
\left(\hat{T}+\int \ddroit\mathbf{r}\,v_{\rm
s}^{\mathbf{w}}(\mathbf{r})\hat{n}(\mathbf{r})\right)\left\vert\Phi^{\mathbf{w}}_K\right\rangle
=\mathcal{E}^{\mathbf{w}}_K\left\vert\Phi^{\mathbf{w}}_K\right\rangle,
\ee
where the ensemble KS potential reads $v_{\rm
s}^{\mathbf{w}}(\mathbf{r})=v_{\rm
ext}(\mathbf{r})+\delta E_{\rm Hxc}^{\mathbf{w}}[n_{\hat{\gamma}_{\rm
s}^{\mathbf{w}}}]/\delta n(\mathbf{r})$ and $0\leq K\leq M$. Note that
the (weight-dependent) KS energy $\mathcal{E}^{\mathbf{w}}_K$ is simply obtained by
summing up the energies of the spin-orbitals that are occupied in
$\Phi^{\mathbf{w}}_K$.\\

From the GOK-DFT ensemble energy expression in
Eq.~(\ref{eq:min_gamma_Ew}) and the expression for the individual
energies in Eq.~(\ref{eq:ind_ener_exp_deriv_wft}), we can now derive
exact density-functional expressions for all the energy levels included
into the ensemble. Indeed, according to the Hellmann--Feynman theorem
and Eq.~(\ref{eq:Hxc_ens_fun}),
we can first express the ensemble energy derivative as follows,
\be
\dfrac{\partial E^{\mathbf{w}}}{\partial
{\tt w}_I}&=&{\rm Tr}\left[{\Delta}\hat{\gamma}_{{\rm s},I}^{\mathbf{w}}
\left(\hat{T}
+\hat{V}_{\rm ext}\right)
\right]
+
\left.\dfrac{\partial E_{\rm xc}^{\mathbf{w}}[n]}{\partial
{\tt w}_I}\right|_{n=n_{\hat{\gamma}_{\rm
s}^{\mathbf{w}}}}
\nonumber\\
&&+\int \ddroit\mathbf{r}\,\dfrac{\delta E_{\rm Hxc}^{\mathbf{w}}[n_{\hat{\gamma}_{\rm
s}^{\mathbf{w}}}]}{\delta n(\mathbf{r})}{\rm
Tr}\left[{{\Delta}\hat{\gamma}_{{\rm
s},I}^{\mathbf{w}}}\hat{n}(\mathbf{r})\right],
\ee      
where ${\Delta}\hat{\gamma}_{{\rm
s},I}^{\mathbf{w}}=\left\vert\Phi^{\mathbf{w}}_I\right\rangle\left\langle\Phi^{\mathbf{w}}_I\right\vert-\left\vert\Phi^{\mathbf{w}}_0\right\rangle\left\langle\Phi^{\mathbf{w}}_0\right\vert$,
thus leading to the following exact expression for the $I$th excitation
energy (see Eq.~(\ref{eq:sc_ks-edft})),
\be\label{eq:final_exp_dEw_over_dwI}
\dfrac{\partial E^{\mathbf{w}}}{\partial
{\tt w}_I}=E_I-E_0=\mathcal{E}^{\mathbf{w}}_I-\mathcal{E}^{\mathbf{w}}_0+
\left.\dfrac{\partial E_{\rm xc}^{\mathbf{w}}[n]}{\partial
{\tt w}_I}\right|_{n=n_{\hat{\gamma}_{\rm
s}^{\mathbf{w}}}},
\ee
which generalizes the GOK-DFT expression for the optical gap~\cite{PRA_GOK_EKSDFT} to
higher excitations. Note that, in the original formulation of GOK-DFT~\cite{PRA_GOK_EKSDFT},
higher excitation energies were obtained from a sequence of
{\it single}-weight-dependent ensemble calculations instead. This is not
necessary anymore here as we use a {\it many}-weight-dependent xc
functional.\\

For formal convenience, we now propose to extend the Levy--Zahariev (LZ)
shift-in-potential procedure~\cite{levy2014ground} to canonical ensembles, in
complete analogy with
Ref.~\cite{senjean2018unified}, 
\be\label{eq:LZ_shift}
\dfrac{\delta E_{\rm
Hxc}^{\mathbf{w}}[n]}{\delta n({\bf r})}
&\rightarrow&\overline{v}^{\mathbf{w}}_{\rm Hxc}[n]({\bf r})=
\dfrac{\delta E_{\rm
Hxc}^{\mathbf{w}}[n]}{\delta n({\bf r})}
\nonumber\\
&&
+
\dfrac{E^{\mathbf{w}}_{\rm Hxc}[n]-\int\ddroit{\bf r}\;
\dfrac{\delta E_{\rm
Hxc}^{\mathbf{w}}[n]}{\delta n({\bf r})}
n({\bf r})}{\int\ddroit{\bf r}\;n({\bf r})}
.
\ee
Thus we obtain the following shifted KS energy expressions,
\be\label{eq:ens_LZ_shifted_KS_ener}
\mathcal{E}^{\mathbf{w}}_K\rightarrow
\overline{\mathcal{E}}^{\mathbf{w}}_K
&=&\mathcal{E}^{\mathbf{w}}_K+
E_{\rm Hxc}^{\mathbf{w}}[n_{\hat{\gamma}_{\rm
s}^{\mathbf{w}}}]
\nonumber\\
&&-
\int\ddroit{\bf r}\;
\dfrac{\delta E_{\rm
Hxc}^{\mathbf{w}}[n_{\hat{\gamma}_{\rm
s}^{\mathbf{w}}}]}{\delta n({\bf r})}
n_{\hat{\gamma}_{\rm
s}^{\mathbf{w}}}({\bf r}).
\ee
As a result [see Eqs.~(\ref{eq:min_gamma_Ew}) and
(\ref{eq:sc_ks-edft})], the exact ensemble energy can be written as a 
weighted sum of shifted KS energies,
\be\label{eq:sum_shifted_KS_ener}
{E}^{\mathbf{w}}=\left(1-\sum^M_{I=1}{\tt
w}_I\right)\overline{\mathcal{E}}^{\mathbf{w}}_0+\sum^M_{I=1}{\tt
w}_I\overline{\mathcal{E}}^{\mathbf{w}}_I.
\ee
Let us stress that, as readily seen from 
Eq.~(\ref{eq:sum_shifted_KS_ener}), the LZ shifting procedure is a way to truly fix (i.e. not anymore up to a
constant) the KS (orbital) energies and, consequently, the ensemble KS
potential. Indeed, as shown in Eq.~(\ref{eq:LZ_shift}), any constant
added to the ensemble Hxc potential will be automatically removed by the
LZ shift. Note also that, by construction, the ensemble Hxc
density-functional energy reads
\be
E^{\mathbf{w}}_{\rm Hxc}[n]=\int\ddroit{\bf
r}\;\overline{v}^{\mathbf{w}}_{\rm Hxc}[n]({\bf r})n({\bf r}).
\ee
\\
As a result, we could think of modeling the shifted Hxc
ensemble potential $\overline{v}^{\mathbf{w}}_{\rm Hxc}[n]({\bf r})$
directly rather than the Hxc ensemble energy, in complete
analogy with Ref.~\cite{levy2014ground}. This is where, in this context, the LZ shift
becomes (much) more than a convenient formal trick. This path will not
be explored further in the rest of the paper and is left for future
work.\\  

Turning finally to the extraction of individual energies, 
we should keep in mind that the (global) LZ shift
does not affect KS energy differences,
\be\label{eq:LZ_shift_noeffect_XE}
\overline{\mathcal{E}}^{\mathbf{w}}_I-\overline{\mathcal{E}}^{\mathbf{w}}_0=\mathcal{E}^{\mathbf{w}}_I-\mathcal{E}^{\mathbf{w}}_0,
\ee 
and therefore, as readily seen from
Eq.~(\ref{eq:final_exp_dEw_over_dwI}), it leaves the true excitation
energies unchanged. It only plays a role in the calculation of exact
energy levels. Indeed, if we combine
Eq.~(\ref{eq:LZ_shift_noeffect_XE}) with Eqs.~(\ref{eq:ind_ener_exp_deriv_wft}),
(\ref{eq:final_exp_dEw_over_dwI}), and (\ref{eq:sum_shifted_KS_ener}),
we obtain the following compact expressions,
\be\label{eq:ind_ener_with_DD}
E_K=\overline{\mathcal{E}}^{\mathbf{w}}_K+\sum^M_{I=1}\left(\delta_{IK}-{\tt
w}_I\right)\left.\dfrac{\partial E_{\rm xc}^{\mathbf{w}}[n]}{\partial
{\tt w}_I}\right|_{n=n_{\hat{\gamma}_{\rm
s}^{\mathbf{w}}}}.
\ee
Once the ensemble xc derivative corrections (second term on the right-hand
side of Eq.~(\ref{eq:ind_ener_with_DD})) have been added to the unshifted KS
energies, applying the LZ shift gives immediately access to {\it any} energy level
in the ensemble, and therefore to any ground- or excited-state molecular property. Unlike in the standard DFT+TD-DFT procedure, a 
single calculation is in principle sufficient.\\

Let us
stress that Eq.~(\ref{eq:ind_ener_with_DD}), which is the key result of
this paper, holds for {\it any} set of ordered ensemble
weights [see Eqs.~(\ref{eq:convex_weights_1}) and
(\ref{eq:convex_weights_0})], including both ground-state ${\tt
w}_{1\leq I\leq M}\rightarrow 0$ and 
equi-ensemble ${\tt w}_{1\leq I\leq M}\rightarrow 1/(M+1)$ limits. In this
respect, it 
generalizes the original formulation of GOK-DFT~\cite{PRA_GOK_EKSDFT} (where
single-weight-dependent xc functionals only were introduced) as well as the
more recent DEC method~\cite{yang2017direct} which, as shown in the following, is recovered
from 
the ground-state limit of Eq.~(\ref{eq:ind_ener_with_DD}). Note finally
that the latter equation extends the recent work of Senjean and
Fromager on charged excitations~\cite{senjean2018unified} 
to neutral excitation processes.\\

\subsection{Connection with existing ensemble
DFT approaches}\label{subsec:connection-standard}

We should point out that our formalism may be connected to the very
recent work of Gould
and Pittalis~\cite{arXiv18_Gould_ens_corr_ener_not_just_sum} on the
expression of density-functional ensemble xc energies in terms of
individual-state contributions. Indeed, starting from
Eq.~(\ref{eq:ind_ener_with_DD}), we could derive, for each state, an individual xc functional that is a bi-functional
of the individual KS density (through the unshifted KS energy) and the
ensemble one. By taking the weighted sum of these bi-functionals we recover a decomposition for the
ensemble xc energy which resembles the one of Gould
and Pittalis~\cite{arXiv18_Gould_ens_corr_ener_not_just_sum}. The connection between the two approaches should
clearly be
explored further. This is left for future work.\\

We also note from Eq.~(\ref{eq:ind_ener_with_DD}) that, even though both
terms on the right-hand side are in principle 
weight-dependent, their sum should of course be
weight-independent. As shown in the following, this will not be the case
anymore when approximate
xc density functionals are used. Note also that, 
in the $\mathbf{w}=
0$ limit, which has been used in previous
works~\cite{PRA_GOK_EKSDFT,PRA_Levy_XE-N-N-1,yang2017direct,sagredo2018can},
the LZ ground-state energy expression
$E_0=\overline{\mathcal{E}}^{\mathbf{w}=0}_0$~\cite{levy2014ground} is recovered and, most importantly, the excited-state
energy expressions can be simplified further as follows,
\be\label{eq:w_zero_limit}
E_J=\overline{\mathcal{E}}^{\mathbf{w}=0}_J+\left.\dfrac{\partial E_{\rm
xc}^{\mathbf{w}}[n_0]}{\partial
{\tt w}_J}\right|_{{\mathbf{w}}=0},
\ee    
where $n_0$ denotes the ground-state density.
As shown in the seminal work of Levy~\cite{PRA_Levy_XE-N-N-1} and readily seen from Eq.~(\ref{eq:w_zero_limit}),
both ground and $J$th excited states cannot be described with the same
KS potential. The latter should indeed exhibit a jump [see the second
term on the right-hand side of Eq.~(\ref{eq:w_zero_limit})], which is
known as the derivative
discontinuity (DD), as the
(neutral) excitation process occurs, exactly like in charged excitation
processes~\cite{senjean2018unified}. If we are able to
model the {\it many}-weight-dependence of the ensemble xc functional,
then we have access to {\it all} ensemble xc derivatives $\partial E_{\rm
xc}^{\mathbf{w}}[n]/\partial{\tt w}_I$ and therefore, by considering
the ground-state $\mathbf{w}\rightarrow 0$ limit, we obtain {\it all}
the DDs.\\

Note finally that, if
we use Eq.~(\ref{eq:w_zero_limit}) to compute the $J$th excitation energy, we
recover the bare KS excitation energy (i.e. the sum of KS orbital energy
differences) to which an ensemble xc derivative correction is applied. When rewritten as follows,     
\be\label{eq:DEC_recovered_w=0}
\left.\dfrac{\partial E_{\rm
xc}^{\mathbf{w}}[n_0]}{\partial
{\tt w}_J}\right|_{{\mathbf{w}}=0}=
\left[\dfrac{\ddroit
E^{\mathbf{w}=\underline{w}^J}_{\rm
xc}[n_0]}{\ddroit w}-
\dfrac{\ddroit
E^{\mathbf{w}=\underline{w}^{J-1}}_{\rm
xc}[n_0]}{\ddroit w}
\right]_{w=0},
\ee
where $\underline{w}^J$ is the ensemble weight vector defined by ${\tt
w}_I=w$ for $1\leq I\leq J$ and ${\tt
w}_I=0$ for $J < I\leq M$, it becomes clear that, in the ground-state
limit, our approach reduces to the DEC one~\cite{yang2017direct}. By
considering a {\it many}-weight-dependent xc functional, we simply extend the applicability of DEC to any kind of ensemble (including
equi-ensembles). We also obtain all the energies from a single ensemble calculation. 
\section{Application to the Hubbard dimer}\label{sec:Hubb_dimer_theory}

We present in the following an implementation of
Eq.~(\ref{eq:ind_ener_with_DD}) for a three-state singlet ensemble. In
the latter case, the convexity conditions in Eqs.~(\ref{eq:convex_weights_1}) and
(\ref{eq:convex_weights_0}) become 
\be 
0\leq {\tt w}_2\leq {1}/{3}
\ee
and
\be
{\tt
w}_2\leq{\tt w}_1\leq (1-{\tt
w}_2)/2.
\ee
The theory is applied to the (not necessarily symmetric) Hubbard
dimer~\cite{carrascal2015hubbard,carrascal2016corrigendum}. It is a simple but non-trivial toy system that is nowadays routinely
used for
exploring new concepts in
DFT~\cite{carrascal2015hubbard,carrascal2016corrigendum,li2018density,carrascal2018linear,sagredo2018can,PRB18_Carsten_dft_non-coll_spin_Hubbard_dimer,senjean2018unified,deur2017exact,deur2018exploring}.
Within this model, the Hamiltonian is simplified as follows (we write operators in second
quantization):
\be\label{eq:Hamil_Hubbard_dimer_model}
\hat{T} &\rightarrow& -t
\sum_{\sigma=\uparrow\downarrow}(\hat{c}^\dagger_{0\sigma}\hat{c}_{1\sigma} +
\hat{c}^\dagger_{1\sigma}\hat{c}_{0\sigma}),\hspace{0.2cm} \hat{W}_{\rm ee}\rightarrow
U\sum^1_{i=0}\hat{n}_{i\uparrow}\hat{n}_{i\downarrow},
\nonumber\\
\hat{V}_{\rm ext}&\rightarrow&\Delta v_{\rm ext}(\hat{n}_1 -
\hat{n}_0)/2,\hspace{0.3cm}\hat{n}_{i\sigma}=\hat{c}^\dagger_{i\sigma}\hat{c}_{i\sigma},
\ee
where $\hat{n}_i=\sum_{\sigma=\uparrow\downarrow}\hat{n}_{i\sigma}$ is
the density operator on site $i$ ($i=0,1$). Note that the external potential reduces to a      single number $\Delta v_{\rm
ext}$ which controls the asymmetry of the model.
The density also reduces
to a single number $n=n_0$ which is the occupation of site 0, given that
$n_1 = 2-n$ (we consider 2-electron canonical ensembles only in this
work).\\

The bi-ensemble consisting of the ground and first singlet excited 
states has been extensively studied in
Refs.~\cite{deur2017exact,deur2018exploring}. Very recently, Sagredo and
Burke~\cite{sagredo2018can} added one more (doubly excited) singlet state 
to the ensemble. As proven in Appendix~\ref{appendix:reduction_tri-to_bi}, the
tri-ensemble analog of the Hohenberg--Kohn functional can be expressed in
terms of the bi-ensemble one.  
As a result, both the ensemble non-interacting kinetic energy $T_{\rm
s}^{\mathbf{w}}(n)$ and the ensemble exact exchange (EEXX) one
$E^{\mathbf{w}}_{\rm x}(n)$ [here $\mathbf{w}\equiv
({\tt w}_1,{\tt w}_2)$] can be determined from their bi-ensemble analogs (see Eqs.~(57) and (62) in Ref.~\cite{deur2017exact}), thus leading to the
simple expressions 
\be\label{eq:Ts_ens_3states}
T_{\rm s}^{\mathbf{w}}(n)=-2t\sqrt{(1-{\tt
w}_1-2{\tt w}_2)^2-(1-n)^2},
\ee
and
\be\label{eq:final_expression_3-state_EEXX}
E^{\mathbf{w}}_{\rm x}(n)
&=&
\dfrac{U}{2}\left[1+{\tt w}_1-\dfrac{(3{\tt w}_1-1)(1-n)^2}{(1-{\tt
w}_1-2{\tt w}_2)^2}\right]
\nonumber
\\
&&-E_{\rm
H}(n),
\ee
where the Hartree energy reads $E_{\rm H}(n)=
U\left(1 + (n - 1)^2\right)$~\cite{deur2017exact}. 
The tri-ensemble density-functional correlation energy is then obtained as follows (
see Appendix~\ref{appendix:reduction_tri-to_bi}), 
\be\label{eq:scaling_relation_tri-ens-corr}
E^{\mathbf{w}}_{\rm c}(n)&=&(1-3{\tt w}_2)E_{\rm c}^{w
}(\nu),
\ee
where $E_{\rm c}^{w}(\nu)$ a bi-ensemble correlation energy, with {\it
effective} weight 
$w=({{\tt w}_1-{\tt
w}_2})/({1-3{\tt w}_2})$ and density $\nu=(n-3{\tt w}_2)/(1-3{\tt
w}_2)$, which can be computed to arbitrary accuracy by Lieb
maximization~\cite{deur2017exact}. The tri- to bi-ensemble
reduction in Eq.~(\ref{eq:scaling_relation_tri-ens-corr}) is of course
{\it not} a general result. It only applies to the Hubbard dimer and originates
from the fact that, in this system, the three singlet energies 
sum up to $2U$ (see Eq.~\ref{eq:simplification_tri-to_bi}).\\

Turning to the non-interacting KS system
with potential $\Delta v_{\rm s}^{{\mathbf{w}}}$, the
(unshifted) energies of the ground-, singly- and doubly-excited states
read $\mathcal{E}^{\mathbf{w}}_0=2\varepsilon_{\rm H}\left(\Delta
v_{\rm s}^{{\mathbf{w}}}\right)$, $\mathcal{E}^{\mathbf{w}}_1=0$, and
$\mathcal{E}^{\mathbf{w}}_2=-2\varepsilon_{\rm H}\left(\Delta
v_{\rm s}^{{\mathbf{w}}}\right)$, respectively, where $\varepsilon_{\rm H}(\Delta v)= 
- \sqrt{t^2 + (\Delta v^2 / 4)}$~\cite{deur2017exact}. Note that the density-functional KS potential
can be simply calculated as $\Delta v_{\rm
s}^{{\mathbf{w}}}(n)=\partial T_{\rm
s}^{\mathbf{w}}(n)/\partial n$~\cite{deur2017exact}. The Hxc potential, which is needed
in the LZ shift-in-potential procedure (see
Eq.~(\ref{eq:LZ_shift})), is then determined as follows, $\Delta v_{\rm
Hxc}^{{\mathbf{w}}}(n)=\Delta v_{\rm s}^{{\mathbf{w}}}(n)-\Delta v_{\rm
ext}$, where $n$ is
the physical tri-ensemble density obtained from the Hamiltonian in
Eq.~(\ref{eq:Hamil_Hubbard_dimer_model}). As shown in Appendix~\ref{appendix:symmetric_dimer}, in the symmetric
case ($\Delta v_{\rm
ext}=0$), the full problem can be solved analytically.\\

\section{Results and discussion}\label{sec:results-perspec}

We have shown in Sec.~\ref{subsec:ind_ener_gok-dft} that individual energy levels can be
extracted, in principle exactly, from a single many-weight-dependent ensemble GOK-DFT calculation by
adding to each (ground- and excited-state) KS energy 
a global LZ-type shift {\it and} an ensemble-based state-specific xc
derivative
correction (see Eqs.~(\ref{eq:ens_LZ_shifted_KS_ener}) and (\ref{eq:ind_ener_with_DD})). In order to assess
the importance of both corrections, we first investigate the deviation
of the KS energies from the exact physical ones. The former are
simply obtained  
by summing up (unshifted) KS orbital energies. Note that, in
contrast to the LZ-shifted ones, these energies are not uniquely
defined because the KS potential is unique {up to an arbitrary
constant}. In the Hubbard dimer model, the latter is chosen such that the
potential sums to zero over the two sites (see
Eq.~(\ref{eq:Hamil_Hubbard_dimer_model})).
\begin{figure}
\resizebox{0.45\textwidth}{!}{
\includegraphics[scale=1.0]{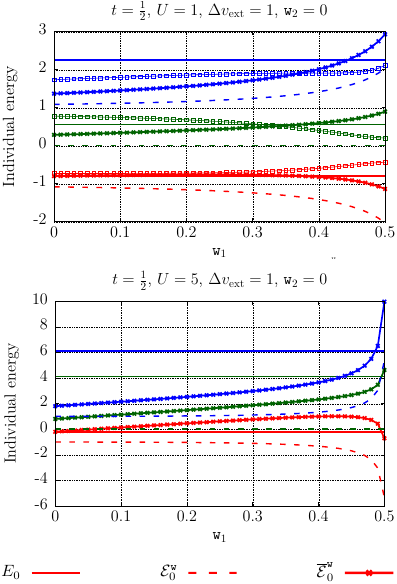}
}
\caption{
Unshifted $\mathcal{E}^{\mathbf{w}}_{K=0,1,2}$ and LZ-shifted
$\overline{\mathcal{E}}^{\mathbf{w}}_{K=0,1,2}$ KS energies obtained for
the asymmetric Hubbard dimer by
varying the first ensemble weight while fixing the second one to
zero. Results are shown for $U/t=2$ (top panel) and $U/t=10$ (bottom
panel). Comparison is made with the
exact energies $E_{K=0,1,2}$. First and second excited-state energies
are shown in green and blue, respectively. EEXX-only results (including both LZ shift
and ensemble xc derivative corrections) are also plotted in the top panel (with squares) for
analysis
purposes.
}
\label{fig:unshift-LZ-shift_ener_asym_U1U5}
\end{figure}
\begin{figure}
\centering
\resizebox{0.45\textwidth}{!}{
\includegraphics[scale=1.0]{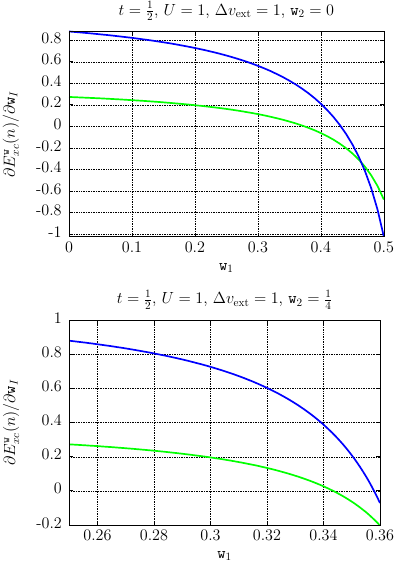}
}
\caption{
Exact ensemble xc derivatives obtained from Eqs.~(\ref{eq:Xener}) and (\ref{eq:final_exp_dEw_over_dwI}) for the asymmetric ($\Delta
v_{\rm ext}/t=2$) Hubbard dimer
with $U/t=2$. First- ($I=1$) and
second-excited-state ($I=2$) derivatives are calculated for the true
(weight-dependent) ensemble density generated from $\Delta v_{\rm
ext}$ and are plotted (in green and blue,
respectively) as functions of the first ensemble weight ${\tt
w}_1$ for ${\tt
w}_2=0$ (top panel) and ${\tt
w}_2=1/4$ (bottom panel).}
\label{fig:each_exact_DDs_U1_deltav1}
\end{figure}
As illustrated in
Fig.~\ref{fig:unshift-LZ-shift_ener_asym_U1U5}, the unshifted KS energies
are found to be substantially lower than the exact energies. It is particularly
striking for the first excited state whose unshifted KS energy 
equals zero, by construction (see Sec.~\ref{sec:Hubb_dimer_theory}). In the
symmetric case (see the Appendix~\ref{appendix:symmetric_dimer} and the supplementary material), the
ground- and second-excited-state unshifted KS energies are equal to $-2t$ and $2t$, respectively.
As a result, varying the ensemble weights has no impact. The situation
is different in the asymmetric case since the unshifted energies can vary with
the weights through the density-functional KS potential. The second (doubly-) excited-state
energy can for example be substantially improved when increasing the
weights. However, the ground-state energy 
deteriorates in that case.\\      

If we now apply the (weight-dependent) LZ shift, more accurate energies
are obtained, as shown in Fig.~\ref{fig:unshift-LZ-shift_ener_asym_U1U5}. Note that, by
construction, the LZ-shifted KS ground-state energy is exact 
when ${\tt w}_1={\tt w}_2=0$. It is important to notice that,
unlike the exact energies, the LZ-shifted ones are (sometimes strongly)
weight-dependent, thus illustrating the importance of modeling 
ensemble xc derivative corrections. The latter are plotted in
Fig.~\ref{fig:each_exact_DDs_U1_deltav1}.
Interestingly, both first- and
second-excited-state derivatives are non-negligible and will therefore contribute to the exact
ground-state energy away from the $\mathbf{w}=0$ limit (see the
second term on the right-hand side of Eq.~(\ref{eq:ind_ener_with_DD})).
Note also that these derivatives are strongly state-dependent. In the
asymmetry and correlation regimes considered in
Fig.~\ref{fig:each_exact_DDs_U1_deltav1}, each derivative vanish for 
particular (state-dependent) weight values. In this
case, the corresponding excitation energy is exactly equal to the KS
one (see Eq.~(\ref{eq:final_exp_dEw_over_dwI})).\\
 
Returning to the LZ-shifted KS energies, their weight dependence becomes even
more important in stronger correlation regimes, as shown in the bottom
panel of
Fig.~\ref{fig:unshift-LZ-shift_ener_asym_U1U5}. 
Note that, in this case, the first and
second excited
states are single- and double-charge transfer states, respectively~\cite{deur2017exact}.
Note also that the first-weight-dependence of the shifted energies is sensitive to
the value of the second weight, as shown in the
supplementary material.
Interestingly, increasing the ensemble weights can
provide more accurate excited-state LZ-shifted energies, often at the
expense of deteriorated ground-state energies. As shown in
Fig.~\ref{fig:unshift-LZ-shift_ener_asym_U1U5} (see also the
supplementary material), this is a general trend that can be seen in all correlation regimes.\\ 

\begin{figure}
\centering
\resizebox{0.45\textwidth}{!}{
\includegraphics[scale=1.0]{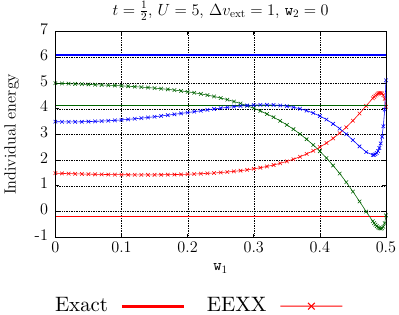}
}
\caption{Ground- (red), first- (green) and second- (blue) excited-state
energies (exact and EEXX-only) plotted as functions of the first
ensemble weight (with ${\tt w}_2=0$) for the asymmetric ($\Delta
v_{\rm ext}/t=2$) Hubbard
dimer with $U/t=10$.}
\label{fig:EEXX_U5}
\end{figure}

Finally, in order to assess the importance of correlation effects in the
calculation of individual energy levels, 
we computed EEXX-only 
LZ shift and ensemble derivative corrections to the unshifted KS energies. Since we used
exact densities (and therefore exact KS potentials), the LZ shift has
been computed with the full
(exact) Hxc
potential in conjunction with the EEXX energy, for the sake of
consistency. In the moderately correlated $U/t=2$ regime (see
the top panel of Fig.~\ref{fig:unshift-LZ-shift_ener_asym_U1U5}),
relatively good total energies
are obtained, which is in agreement with the DEC/EEXX results of Ref.~\cite{sagredo2018can}. 
Interestingly, the doubly-excited state energy is the one that exhibits the weakest
weight dependence. As shown in Fig.~\ref{fig:EEXX_U5}, in the $\Delta v_{\rm
ext}/t=2$ asymmetry regime, EEXX fails
dramatically for the larger $U/t=10$ value. Total energies become strongly weight-dependent and
their ordering is wrong for a wide range of weight values. The latter
observation was actually expected for small weight values on the basis
of 
Ref.~\cite{deur2017exact} (where we see in Fig. 1 that, for $2t=1$, $U=5$ and
$\Delta v_{\rm ext}=1$, the ground-state density is close to 1, which
corresponds to the symmetric case) and
Appendix~\ref{appendix:symmetric_dimer}, where the EEXX
energies are derived for the symmetric Hubbard dimer (see
Eq.~(\ref{eq:symmetric_EEXX_ener})).   \\ 

\section{Single versus sequence of ensemble calculations
}\label{sec:compar_gokII}

While, in conventional GOK-DFT approaches, excitation (or individual) energies are
extracted from a sequence of ensemble calculations (where ensemble
weights are controlled by a single one $w$), we have shown in this work
that a single ensemble calculation is sufficient provided, of course,
that the {\it many}-weight-dependence of the ensemble xc functional is
known. The two approaches are equivalent in the exact theory
but they may give different results when density-functional approximations
are used. This is analyzed further in the rest of this section at the
EEXX level of approximation.\\
 
Let us first rewrite the exact individual energy expressions
within the GOKII approach~\cite{yang2017direct} (see
Eq.~(\ref{eq:gokII_weights})) where both bi- and
tri-ensemble calculations are needed for extracting the three lowest
energies. From the bi-ensemble energy 
\be
E^{(w,0)}=(1-w)E_0+wE_1,
\ee
we can extract both ground- and first-excited-state energies as follows,
\be\label{eq:bi-ensemble-E0E1}
E_0&=&E^{(w,0)}-w\dfrac{dE^{(w,0)}}{dw},
\nonumber\\
E_1&=&E^{(w,0)}+(1-w)\dfrac{dE^{(w,0)}}{dw},
\ee
which is equivalent (for these two states) to a tri-ensemble
calculation where ${\tt w}_1=w$ and ${\tt
w}_2=0$. On the other hand, we have the tri-ensemble energy (with ${\tt w}_1={\tt
w}_2=w$),
\be
E^{(w,w)}=(1-2w)E_0+wE_1+wE_2,
\ee
from which we can extract, when combined with the bi-ensemble one, the
second-excited-state energy: 
\be\label{eq:2nd_ex_state_gokII}
E_2&=&E_0-(E_1-E_0)+\dfrac{dE^{(w,w)}}{dw}
\nonumber\\
&=&E^{(w,0)}-(1+w)\dfrac{dE^{(w,0)}}{dw}
+\dfrac{dE^{(w,w)}}{dw}.
\ee
If, like in Sec.~\ref{sec:Hubb_dimer_theory}, we use a single
ensemble calculation instead (with ${\tt w}_1={\tt
w}_2=w$ for ease of comparison), then  
individual energies will be determined as follows (see Eq.~(\ref{eq:ind_ener_exp_deriv_wft})),
\be\label{eq:E0-E2_tri-ens}
E_0&=&E^{(w,w)}-w\left.\dfrac{\partial E^{(w,{\tt w}_2)}}{\partial w}\right|_{{\tt w}_2=w}
-w\left.\dfrac{\partial E^{({\tt w}_1,w)}}{\partial w}\right|_{{\tt w}_1=w}
\nonumber\\
&=&E^{(w,w)}-w\dfrac{dE^{(w,w)}}{dw},
\ee 
and
\be\label{eq:E0-E2_tri-ens_excited_states}
E_1&=&E_0+\left.\dfrac{\partial E^{(w,{\tt w}_2)}}{\partial w}\right|_{{\tt w}_2=w},
\nonumber\\
E_2&=&E_0+\left.\dfrac{\partial E^{({\tt w}_1,w)}}{\partial
w}\right|_{{\tt w}_1=w}.
\ee
Note that we use the latter expressions rather than the (equivalent)
ones in Eq.~(\ref{eq:ind_ener_with_DD}) for ease of comparison.\\

\begin{figure}
\centering
\resizebox{0.45\textwidth}{!}{
\includegraphics[scale=1.0]{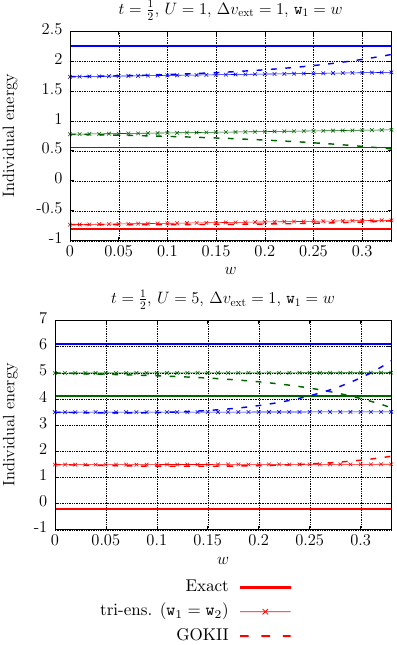}
}
\caption{
Comparison of single tri-ensemble (with ${\tt w}_1={\tt w}_2=w$) and GOKII approaches at the   
EEXX-only level of approximation in the
asymmetric Hubbard dimer for $U/t=2$ (top panel) and $U/t=10$ (bottom panel). Red, green and blue
colors are used for the ground, first and second excited states,
respectively. See text for further details.}
\label{fig:compar_with_gok2}
\end{figure}

As readily seen from
Eqs.~(\ref{eq:bi-ensemble-E0E1})-(\ref{eq:E0-E2_tri-ens}), the two
approaches become identical (and equivalent to DEC~\cite{yang2017direct,sagredo2018can}) in the $w\rightarrow 0$ limit, even when
approximate ensemble energies are used. For larger $w$ values (in the
range $0<w\leq 1/3$), the two methods will 
give substantially different results for the excited states
when the EEXX-only approximation is used, as illustrated
in Fig.~\ref{fig:compar_with_gok2}. While, in our (single-calculation-based)
approach, individual
energies are increasingly insensitive to the value of the tri-ensemble
weight as $U/t$ increases, the GOKII excited-state energies exhibit an
important weight dependence. Interestingly, as $w$ increases, they
become closer to the exact energies. In the large $U/t$ regime (see
the bottom panel of Fig.~\ref{fig:compar_with_gok2}), increasing $w$
restores the correct ordering of the excited states.\\

In the Hubbard dimer, the EEXX-only
individual energies can be expressed as explicit functionals of the
ensemble density (see
Appendix~\ref{appendix:dens_fun_ener_compar_gok2}), thus allowing for a
better understanding of these results. The key difference between GOKII and
the single tri-ensemble calculation approach is the ensemble density
itself. As $U/\Delta
v_{\rm ext}$ and $\Delta
v_{\rm ext}/t$ increase, the tri-ensemble density becomes closer to 1
(see Appendix~\ref{appendix:dens_fun_ener_compar_gok2}), thus explaining why tri-ensemble-based-only energies
are essentially the (weight-independent) ones obtained at the symmetric
EEXX level. Note that, in the latter case, the excited states are wrongly ordered (see
Appendix~\ref{appendix:symmetric_dimer}). On the other hand, the
bi-ensemble density varies as $1+w$ in the same asymmetry and
correlation regime. As a result, analytical expressions can be derived
for the variation in $U$ and $w$ of the  
GOKII/EEXX energies (see Eqs.~(\ref{eq:bi-ens_contr_E2_largeU}) and
(\ref{eq:bi-ens_contr_E0E1_largeU})), thus providing a rationale for the
results shown in Fig.~\ref{fig:compar_with_gok2}. As proven in Appendix~\ref{appendix:dens_fun_ener_compar_gok2}, the improvement of the
second-excited-state energy as $w$ increases is exclusively due to the bi-ensemble
contribution (two first terms on the right-hand side of
Eq.~(\ref{eq:2nd_ex_state_gokII})). The good performance of GOKII/EEXX (in terms of
total excited-state energies) 
may be
specific to the Hubbard dimer. Nevertheless, it clearly shows that
the choice of ensemble and extraction procedure is crucial when using
density-functional approximations.  

\section{Conclusions and perspectives}\label{conclusions}

A generalized many-weight-dependent formulation of GOK-DFT has been 
explored, thus leading to an in-principle-exact energy level extraction
procedure that applies to any (ground or excited) state in the ensemble
and relies on a single ensemble DFT
calculation. The latter consists, like a conventional DFT
calculation, in solving a single set of self-consistent KS equations
where the
orbitals are fractionally occupied (the occupation numbers are determined from the ensemble
weights). The theory has been applied to the Hubbard dimer.
The two corrections that should in principle be added to the bare KS energies (namely the global LZ shift and
a state-specific (ensemble-based) xc
derivative correction) were both shown to be important in the calculation of accurate and
weight-independent energy levels. In order to turn the method
into a practical computational tool, {\it ab initio}
many-weight-dependent xc density-functional approximations should be
developed. This can be achieved, for example, by applying GOK-DFT to finite uniform electron
gases~\cite{JCP17_T2_GLDA_EXX}. A nice feature of such model systems is that
both ground and excited states share the same density which is the
ensemble density itself. Consequently, in this particular case, the
density-functional ensemble xc energy is simply the weighted sum of the
individual-state xc energies. Work is currently in progress in this
direction. Finally, regarding the application of the theory to
photochemical processes, we would like to explore the possibility of
extracting non-adiabatic couplings from a GOK-DFT calculation. It may be
useful, for that
purpose, to extend the theory to the time-dependent linear response
regime. This is left for future work.\\

\section{Supplementary Material}

We provide complementary curves showing the variation in the first
ensemble weight ${\tt w}_1$ of individual energy levels (before and after the LZ shift)
for ${\tt w}_2=0$ or ${\tt
w}_2=1/4$ in various correlation and asymmetry regimes of the Hubbard
dimer.  

\begin{acknowledgments}
The authors thank the ANR (MCFUNEX project, Grant No. ANR-14-CE06-
0014-01) for funding. E.~F. would also like to thank P.~F.~Loos for
stimulating discussions.

\end{acknowledgments}


\appendix

\section{Connection between exact tri- and bi-ensemble
functionals}\label{appendix:reduction_tri-to_bi}

We start
from the Lieb-maximization-based expression for the three-state ensemble analog of
the ($U$- and $t$-dependent) Hohenberg--Kohn (HK) functional which reads in this
context~\cite{deur2017exact},
\be\label{eq:Fw_3states}
F^{\mathbf{w}}(n)=\sup_{\Delta v}\Big\lbrace
&&\left(1-{\tt w}_1-{\tt w}_2\right)E_0(\Delta v)+
{\tt w}_1E_1(\Delta v)
\nonumber\\
&&
+{\tt w}_2E_2(\Delta v)
 - 
{\Delta v}(1 - n)
\Big\rbrace.
\ee
Since the three singlet energies sum up to $2U$ (see Eq.~(26) in
Ref.~\cite{senjean2017local}), the expression in Eq.~(\ref{eq:Fw_3states}) can be
simplified as follows,  
\be\label{eq:simplification_tri-to_bi}
F^{\mathbf{w}}(n)&=&2U{\tt w}_2+\sup_{\Delta v}\Big\lbrace
\left(1-{\tt w}_1-2{\tt w}_2\right)E_0(\Delta v)
\nonumber\\
&&+ 
({\tt w}_1-{\tt w}_2)E_1(\Delta v)
 - 
{\Delta v}(1 - n)\Big\rbrace, 
\ee
which can then be rewritten formally as

\be\label{eq:link_two_three_states_F}
F^{\mathbf{w}}(n)&=&2U{\tt w}_2+(1-3{\tt w}_2)F^w(\nu),
\ee
where $w=({{\tt w}_1-{\tt
w}_2})/({1-3{\tt w}_2})$ and $\nu=(n-3{\tt w}_2)/(1-3{\tt w}_2)$ are
{\it effective} bi-ensemble weight and density, respectively, and the
corresponding bi-ensemble functional reads~\cite{deur2017exact,deur2018exploring}
\be
F^w(\nu)=\sup_{\Delta v}\lbrace&& (1-w)E_0(\Delta v)+
{w}E_1(\Delta v)
\nonumber\\
&&-
{\Delta v}(1 - \nu)
\rbrace.
\ee
From the non-interacting ($U=0$) limit of
Eq.~(\ref{eq:link_two_three_states_F}) and Eq.~(57) in
Ref.~\cite{deur2017exact} we obtain the expression for the tri-ensemble
non-interacting kinetic energy in Eq.~(\ref{eq:Ts_ens_3states}). Since
the Hx energy is the first-order contribution to the Taylor expansion in
$U$ of the ensemble HK
functional~\cite{gould2017hartree,deur2018exploring}, it comes from
Eq.(\ref{eq:link_two_three_states_F}),    
\be
E^{\mathbf{w}}_{\rm x}(n)&=&2U{\tt w}_2+(1-3{\tt w}_2)E_{\rm x}^{w}\left(\nu\right)
\nonumber\\
&&+(1-3{\tt w}_2)E_{\rm H}\left(\nu\right)
-E_{\rm H}(n),
\ee
thus leading, with Eq.~(62) of Ref.~\cite{deur2017exact}, to the
expression in Eq.~(\ref{eq:final_expression_3-state_EEXX}). The
correlation energy corresponds to all higher-order
contributions in $U$ to the HK functional, which leads to the scaling
relation in Eq.~(\ref{eq:scaling_relation_tri-ens-corr}).\\


\section{Symmetric Hubbard dimer}\label{appendix:symmetric_dimer}

In the particular case of a symmetric dimer ($\Delta v_{\rm ext}=0$),
the LZ-shifted KS energies can be simplified as follows, 
$\overline{\mathcal{E}}^{\mathbf{w}}_0=-2t+C^{\mathbf{w}}_{\rm
LZ}$, $\overline{\mathcal{E}}^{\mathbf{w}}_1=C^{\mathbf{w}}_{\rm
LZ}$, and $\overline{\mathcal{E}}^{\mathbf{w}}_2=2t+C^{\mathbf{w}}_{\rm
LZ}$, where the shift equals
\be\label{eq:LZ_shift_sym_case}
C^{\mathbf{w}}_{\rm
LZ}&=&E^{\mathbf{w}}_{\rm Hxc}(n=1)=\dfrac{U}{2}(1+{\tt w}_1)
\nonumber\\
&&+2t(1-2{\tt w}_2-{\tt w}_1)
.\left(1-\sqrt{1+[{U^2}/({16t^2)]}}\right).
\ee  
As readily seen from Eq.~(\ref{eq:LZ_shift_sym_case}) [see also the
plots in
the supplementary material], 
these energies
are weight-dependent, thus illustrating the importance of the ensemble
xc derivative corrections
in the calculation of physical (weight-independent) energies.
Interestingly, if correlation is neglected in both the LZ shift and the
ensemble derivative 
corrections [the approximation is
referred to as EEXX in the text],  
we obtain the following weight-independent energy expressions,
\be\label{eq:symmetric_EEXX_ener}
E^{\rm EEXX}_0=-2t+\dfrac{U}{2},\hspace{0.1cm} E^{\rm EEXX}_1=U,
\hspace{0.1cm}E^{\rm EEXX}_2=2t+\dfrac{U}{2}.
\ee
While EEXX (which can be seen as perturbation theory through first order
in $U/t$) gives the exact energy level for the first (symmetric) excited state in all
correlation regimes, the individual energy levels 
are well described 
for the ground and second excited states only in the
symmetric weakly correlated regime (i.e. for small $U/t$ values). When the correlation is strong, the
excited levels are actually wrongly ordered. Note that, {\it in the symmetric
case}, the
second (double) excitation energy is not affected by the EEXX-only
derivative correction. Indeed, for $n=1$, the EEXX density functional
does not vary with ${\tt w}_2$ (see
Eq.~(\ref{eq:final_expression_3-state_EEXX})) and, therefore, the
second-excited-state ensemble 
derivative is equal to zero. This result was expected on the basis of
the recently published DEC/EEXX results for the Hubbard dimer
(see Eq.~(6) of Ref.~\cite{sagredo2018can}) and Eq.~(\ref{eq:DEC_recovered_w=0}).
\section{Expressions for bi- and tri-ensemble-based EEXX
density-functional energies}\label{appendix:dens_fun_ener_compar_gok2}

In order to derive analytical expressions for the energy levels within the
EEXX approximation, we start from the general ensemble EEXX-only
density-functional energy expression,
\be
E^{\mathbf{w}}_{\rm EEXX}(n)&=&T_{\rm s}^{\mathbf{w}}(n)+E_{\rm
H}(n)+E^{\mathbf{w}}_{\rm x}(n)
\nonumber\\
&&+(1-n)\Delta v_{\rm ext},
\ee
and the corresponding ensemble derivatives,
\be
&&\dfrac{\partial E^{\mathbf{w}}_{\rm EEXX}(n)}{\partial {\tt w}_1}=
\dfrac{2t(1-{\tt
w}_1-2{\tt w}_2)}{\sqrt{(1-{\tt
w}_1-2{\tt w}_2)^2-(1-n)^2}}
\nonumber\\
&&+\dfrac{U}{2}\left[1
-\dfrac{\Big(1-3(2{\tt w}_2-{\tt w}_1)\Big)(1-n)^2}{(1-{\tt
w}_1-2{\tt w}_2)^3}\right]
\ee 

and
\be
&&\dfrac{\partial E^{\mathbf{w}}_{\rm EEXX}(n)}{\partial {\tt w}_2}=
\dfrac{4t(1-{\tt
w}_1-2{\tt w}_2)}{\sqrt{(1-{\tt
w}_1-2{\tt w}_2)^2-(1-n)^2}}
\nonumber\\
&&-\dfrac{2U(3{\tt
w}_1-1)(1-n)^2}{(1-{\tt
w}_1-2{\tt w}_2)^3}
.\ee
The ensemble density-functional energies and derivatives from which we
can extract individual energies within both GOKII and tri-ensemble-only
approaches are 
\be
&&E_{\rm EEXX}^{(w,0)}(n)=-2t\sqrt{(1-{
w})^2-(1-n)^2}
\nonumber\\
&&+\dfrac{U}{2}\left[1+{w}-\dfrac{(3{w}-1)(1-n)^2}{(1-{
w})^2}\right]+(1-n)\Delta v_{\rm ext},
\ee
\be
&&\dfrac{dE_{\rm EEXX}^{(w,0)}(n)}{dw}=
\dfrac{2t(1-{
w})}{\sqrt{(1-{
w})^2-(1-n)^2}}
\nonumber\\
&&+\dfrac{U}{2}\left[1
-\dfrac{(1+3{w})(1-n)^2}{(1-{
w})^3}\right],
\ee
\be
&&E_{\rm EEXX}^{(w,w)}(n)=-2t\sqrt{(1-3{w})^2-(1-n)^2}
\nonumber\\
&&+\dfrac{U}{2}\left[1+{w}-\dfrac{(1-n)^2}{(3w-1)}\right]+(1-n)\Delta
v_{\rm ext},
\ee
\be
\left.\dfrac{\partial E_{\rm EEXX}^{(w,{\tt w}_2)}(n)}{\partial w}\right|_{{\tt
w}_2=w}&=&
\dfrac{2t(1-3{
w})}{\sqrt{(1-3{
w})^2-(1-n)^2}}
\nonumber\\
&&+\dfrac{U}{2}\left[1
-\dfrac{(1-n)^2}{(1-3{
w})^2}\right],
\ee
and
\be
\left.\dfrac{\partial E_{\rm EEXX}^{({\tt w}_1,w)}(n)}{\partial w}\right|_{{\tt
w}_1=w}&=&\dfrac{4t(1-3{
w})}{\sqrt{(1-3{
w})^2-(1-n)^2}}
\nonumber\\
&&+2U\dfrac{(1-n)^2}{(1-3{
w})^2}.
\ee
At the GOKII/EEXX level, the energies are approximated as follows,
\be\label{eq:ind_ener_dens_GOKII}
E_0&\approx& E^{{\rm
EEXX}}_0\left(n^{(w,0)}\right), 
\nonumber\\
E_1&\approx& E^{{\rm
EEXX}}_1\left(n^{(w,0)}\right),
\nonumber\\
E_2&\approx& E^{{\rm
EEXX}(b)}_2\left(n^{(w,0)}\right)+E^{{\rm
EEXX}(t)}_2\left(n^{(w,w)}\right),
\ee
where the (physical) bi- and tri-ensemble densities can be written as 
\be
n^{(w,0)}=(1-w)n^0_{\Delta v_{\rm ext}}+w\,n^1_{\Delta v_{\rm ext}}
\ee
and 
\be\label{eq:true_tri-ens_density}
n^{(w,w)}&=&(1-2w)n^0_{\Delta v_{\rm ext}}+w\,n^1_{\Delta v_{\rm
ext}}+w\,n^2_{\Delta v_{\rm ext}}
\nonumber\\
&=&3w+(1-3w)n^0_{\Delta v_{\rm ext}},
\ee
respectively. Note that, in Eq.~(\ref{eq:true_tri-ens_density}), we used the fact that the three singlet
densities (which are obtained by differentiating the energies with
respect to the external potential~\cite{deur2017exact}) sum up to 3, as a consequence of the fact that the energies
sum up to $2U$ (which does not depend on the external potential).
The density-functional ground- and first-excited-state energies in Eq.~(\ref{eq:ind_ener_dens_GOKII}) are 
\be
E^{{\rm
EEXX}}_0\left(n\right)&=&E_{\rm EEXX}^{(w,0)}(n)-w\dfrac{dE_{\rm
EEXX}^{(w,0)}(n)}{dw}
,
\nonumber\\
E^{{\rm
EEXX}}_1\left(n\right)&=&E_{\rm EEXX}^{(w,0)}(n)+(1-w)\dfrac{dE_{\rm
EEXX}^{(w,0)}(n)}{dw},
\ee
while the bi- and tri-ensemble contributions to the second-excited-state
energy (see Eq.~(\ref{eq:2nd_ex_state_gokII})) are
\be
E^{{\rm
EEXX}(b)}_2\left(n\right)=E_{\rm EEXX}^{(w,0)}(n)-(1+w)\dfrac{dE_{\rm
EEXX}^{(w,0)}(n)}{dw},
\nonumber\\
\ee
and
\be
E^{{\rm
EEXX}(t)}_2\left(n\right)&=&
\dfrac{6t(1-3{
w})}{\sqrt{(1-3{
w})^2-(1-n)^2}}
\nonumber\\
&&+\dfrac{U}{2}\left[1
+3\dfrac{(1-n)^2}{(1-3{
w})^2}\right],
\ee
respectively.
Note that, in the symmetric case, the three energies obtained
from Eq.~(\ref{eq:ind_ener_dens_GOKII}) are $-2t+(U/2)$, $U$, and
$2t+(U/2)$, which is exactly what is obtained when performing a single
tri-ensemble calculation (see Appendix~\ref{appendix:symmetric_dimer}).\\
    
In the particular case where $t<<\Delta v_{\rm ext}<<U$, we have
$n^0_{\Delta v_{\rm ext}}\approx1$ and
$n^1_{\Delta v_{\rm ext}}\approx 2$ (see Fig.~1 in
Ref.~ \cite{deur2017exact}). The bi- and tri-ensemble densities are
then equal to
$n^{(w,0)}=1+w$ and $n^{(w,w)}=1$, respectively. Consequently, the
tri-ensemble contribution to the second-excited-state energy becomes 
weight-independent and equal to $6t+(U/2)$ while the bi-ensemble
contribution varies in $w$ as follows,  
\be\label{eq:bi-ens_contr_E2_largeU}
\left.E^{{\rm
EEXX}(b)}_2\left(n\right)\right|_{n=1+w}
\approx
\dfrac{Uw^2\left(1+3w^2\right)}{(1-w)^3}
-w\Delta v_{\rm ext}.
\nonumber\\
\ee
We can show similarly that the ground- and first-excited-state energies vary in
$w$ as follows,
\be\label{eq:bi-ens_contr_E0E1_largeU}
\left.E^{{\rm
EEXX}}_0\left(n\right)\right|_{n=1+w}
&\approx&
\dfrac{U}{2}+\dfrac{Uw^2[1+3w(2w-1)]}{2(1-w)^3}
-w\Delta v_{\rm ext},
\nonumber\\
\left.E^{{\rm
EEXX}}_1\left(n\right)\right|_{n=1+w}
&\approx&
U-\dfrac{3Uw^3}{(1-w)^2}-w\Delta v_{\rm ext}.
\ee

Turning to the single tri-ensemble calculation approach, individual energies can
be approximated as follows at the EEXX level,
\be
E_K\approx \mathcal{E}^{{\rm
EEXX}}_K\left(n^{(w,w)}\right), 
\ee
where, according to Eqs.~(\ref{eq:E0-E2_tri-ens}) and
(\ref{eq:E0-E2_tri-ens_excited_states}), 
\be
&&\mathcal{E}^{{\rm
EEXX}}_0\left(n\right)=E_{\rm EEXX}^{(w,w)}(n)
\nonumber\\
&&-w\left(
\left.\dfrac{\partial E_{\rm EEXX}^{(w,{\tt w}_2)}(n)}{\partial w}\right|_{{\tt
w}_2=w}
+
\left.\dfrac{\partial E_{\rm EEXX}^{({\tt w}_1,w)}(n)}{\partial w}\right|_{{\tt
w}_1=w}
\right),
\ee 
\be
&&\mathcal{E}^{{\rm
EEXX}}_1\left(n\right)=\mathcal{E}^{{\rm
EEXX}}_0\left(n\right)
+
\left.\dfrac{\partial E_{\rm EEXX}^{(w,{\tt w}_2)}(n)}{\partial w}\right|_{{\tt
w}_2=w}
,
\ee
and
\be
&&\mathcal{E}^{{\rm
EEXX}}_2\left(n\right)=\mathcal{E}^{{\rm
EEXX}}_0\left(n\right)
+
\left.\dfrac{\partial E_{\rm EEXX}^{({\tt w}_1,w)}(n)}{\partial w}\right|_{{\tt
w}_1=w}.
\ee




\newcommand{\Aa}[0]{Aa}

\end{document}